\documentclass[hyper, notoc]{JHEP3} 
\usepackage{epsfig}   
  
\newcommand{\be}{\begin{equation}}  
\newcommand{\ee}{\end{equation}}  
\newcommand{\bea}{\begin{eqnarray}}     
\newcommand{\eea}{\end{eqnarray}}

\title{Running Into New Territory in SUSY Parameter Space}   
\author{Puneet Batra \\ Department of Physics, Stanford University, Stanford, CA 94305}   
\author{Antonio Delgado, David E. Kaplan\\Department of Physics and Astronomy, Johns  
Hopkins University, Baltimore, MD 21218}  
\author{Tim M.P. Tait \\ Fermi National Accelerator Laboratory, Batavia, IL 60510}  
 
\abstract{The LEP-II bound on the light Higgs mass rules out the vast
majority of parameter space left to the Minimal Supersymmetric
Standard Model (MSSM) with weak-scale soft-masses. This suggests the
importance of exploring extensions of the MSSM with non-minimal Higgs
physics.  In this article, we explore a theory with an additional
singlet superfield and an extended gauge sector.  The theory has a
number of novel features compared to both the MSSM and Next-to-MSSM,
including easily realizing a light CP-even Higgs mass consistent with
LEP-II limits, $\tan \beta \lesssim 1$, and a lightest Higgs which is
charged.  These features are achieved while remaining consistent with
perturbative unification and without large stop-masses. Discovery
modes at the Tevatron and LHC are discussed.}
 
\preprint{FERMILAB-PUB-04/038-T\\ SU-ITP-04/12}  
  
\begin{document}  
  
\section{Introduction}  
 
\label{sec:intro} 
 
The minimal supersymmetric standard model (MSSM) is 
the most-studied theory of new physics at the weak-scale. 
Nevertheless, extensions to the MSSM 
are profligate; considerable effort has been applied, for example, to 
address shortcomings in the Higgs sector ($H, \bar{H}$) of the MSSM. 
In particular, the MSSM allows for a superpotential mass term $\mu H \bar{H}$   and fixes the quartic-coupling to the size of the (relatively small) 
electroweak gauge couplings. The unexplained coincidence of scale 
between $\mu$ and the soft-masses is the $\mu$-problem. The small size 
of the quartic predicts at tree-level a lightest CP-even Higgs state, 
$h^0$, with mass less than $M_Z$---a result ruled out by searches at 
LEP-II \cite{unknown:2001xx}.  
 
Additional quantum corrections from the top sector can raise $m_{h^0}$ 
to $\sim$130 GeV---though only in the case of large $\tan\beta$ (the 
ratio of the vacuum expectation values (VEV's) of the two Higgses), 
TeV stop masses, and a maximal stop mixing angle 
\cite{Carena:1995wu}. Although heavy scalar tops are not precluded by 
any experiment, they dominate the fine-tuning in the electroweak 
sector through 1-loop contributions to the Higgs doublet soft-masses; 
the cost of opening up parameter space in the MSSM is to reintroduce 
fine-tuning into the Higgs potential. 
 
A possible remedy to both of these short-comings is the Next-to-Minimal 
Supersymmetric Standard Model (NMSSM)  
\cite{Ellis:1988er}, in which  
a gauge singlet superfield ($S$) with 
superpotential interaction $\lambda_S S H \bar{H}$ is added to the 
MSSM. The usual $\mu$ term is prevented by imposing a $Z_3$ symmetry. 
$S$ receives a VEV proportional to its 
soft-mass, leading to the desired soft-scale $\mu$ term $\lambda_S 
\langle S \rangle H \bar{H}$. In this case, the admixture of the 
singlet into the electroweak symmetry breaking sector can ease 
experimental constraints even though light CP-even states still exist 
\cite{Miller:2003ay}.  Alternatively, as 
noted in \cite{Haber:1986gz,Espinosa:1998re},  
if $S$ has a large soft-mass and is integrated 
out in the non-supersymmetric limit, then the  effective theory contains a 
new quartic-coupling $|\lambda_S|^2 |H \bar{H}|^2$ which independently 
lifts the lightest CP-even state at tree level---irrespective of the 
stop masses.  Although one gives up using $S$ as a solution to the  
$\mu$-problem, one can pursue additional options such as the Giudice-Masiero mechanism \cite{Giudice:1988yz}.  
 
The models described in \cite{Espinosa:1998re} which attempt to raise
the CP-even states without heavy stops are limited in their effect on
the lightest Higgs mass by requiring perturbativity of all couplings
up the GUT scale.  The issue is that every component of the Higgs
quartic coupling is infrared free.  The requirement that $\lambda_S$
remains perturbative to the GUT scale forces one to choose $\lambda_S
\lesssim 0.6$ at the weak scale, corresponding to $m_{h^0} \lesssim
160$ GeV. Significantly larger values of $\lambda_S$ (and hence
$m_{h^0}$) can be had at the weak-scale if one gives up standard
perturbative unification and simply cuts-off the (now assumed)
effective theory at the scale where $\lambda_S$ hits a Landau pole
\cite{Tobe:2002zj}.

In \cite{Batra:2003nj} these perturbative constraints led us to extend
the MSSM gauge sector by a non-Abelian gauge group which adds an
asymptotically-free contribution to the quartic coupling. The extended
gauge structure allowed us to consider significantly larger quartics
at the weak scale without spoiling GUT unification. After imposing
constraints from precision electroweak observables and requiring no
fine-tuning in the Higgs sector, the resultant increase in the
lightest CP-even mass bound was dramatic: $m_{h^0} \lesssim 350$
GeV. Similar bounds were found recently in models which use hard
effects from low-scale SUSY breaking \cite{Casas:2003jx}, and in
models which interpret the Landau pole in the quartic as a
compositeness scale \cite{Harnik:2003rs}.  

In this article we explore an alternative which combines the benefits
of the MSSM-plus-singlet theories with those of the gauge-extension
models.  We raise $m_{h^0}$ by including a singlet coupling $\lambda_S
S H \bar{H}$, and as in \cite{Espinosa:1998re}, we integrate $S$ out
by giving it a relatively large soft-mass ($\sim$ 1 TeV).  In the
absence of further ingredients, this would leave us with the limited
increase in $m_{h^0}$ corresponding to $\lambda_S \lesssim 0.6$ from
the perturbative unification bound mentioned above.  The bound is
removed when we include an asymptotically-free $SU(2)$ interaction
which counteracts the tendency of $\lambda_S$ to drive itself large at
high scales. Atypical regions of supersymmetric parameter space,
specifically those with small $\tan{\beta}\  \& \ m_{H^+} < m_t, m_{h^0}$,
are motivated by our construction. Regions with light $m_{H^+}$ have
previously been identified in other MSSM extensions
\cite{Panagiotakopoulos:2000wp}, and such regions might prove more
generic as the exploration of models beyond-the-MSSM continues.

The resulting model (Section \ref{sec:model}) is structurally similar
to the that of Ref.~\cite{Batra:2003nj}, although we focus on the
effect of the new singlet interaction and do not take advantage of the
additional $D$-term contributions to the Higgs quartic as was done in
\cite{Batra:2003nj}.  Section \ref{sec:constraints} describes how the
extended gauge sector helps to keep both $\lambda_S$ and the top
Yukawa interaction $y_{t}$ (including regions where $\tan{\beta} < 1$)
under perturbative control and consistent with GUT-scale
gauge-coupling unification. Readers interested in the generic
properties of the allowed parameter space should skip to Section
\ref{sec:pheno}, where we find not only that the upper bound on the
lightest CP-even state is 250 GeV, but also that the lightest state in
the Higgs spectrum can be $H^{\pm}$ and discuss the novel
phenomenology.  Section~\ref{sec:conclusions} contains our
conclusions.
 
\section{The Gauge-Extended MSSM + Singlet} 
\label{sec:model} 
 
\FIGURE[r]{\epsfig{file=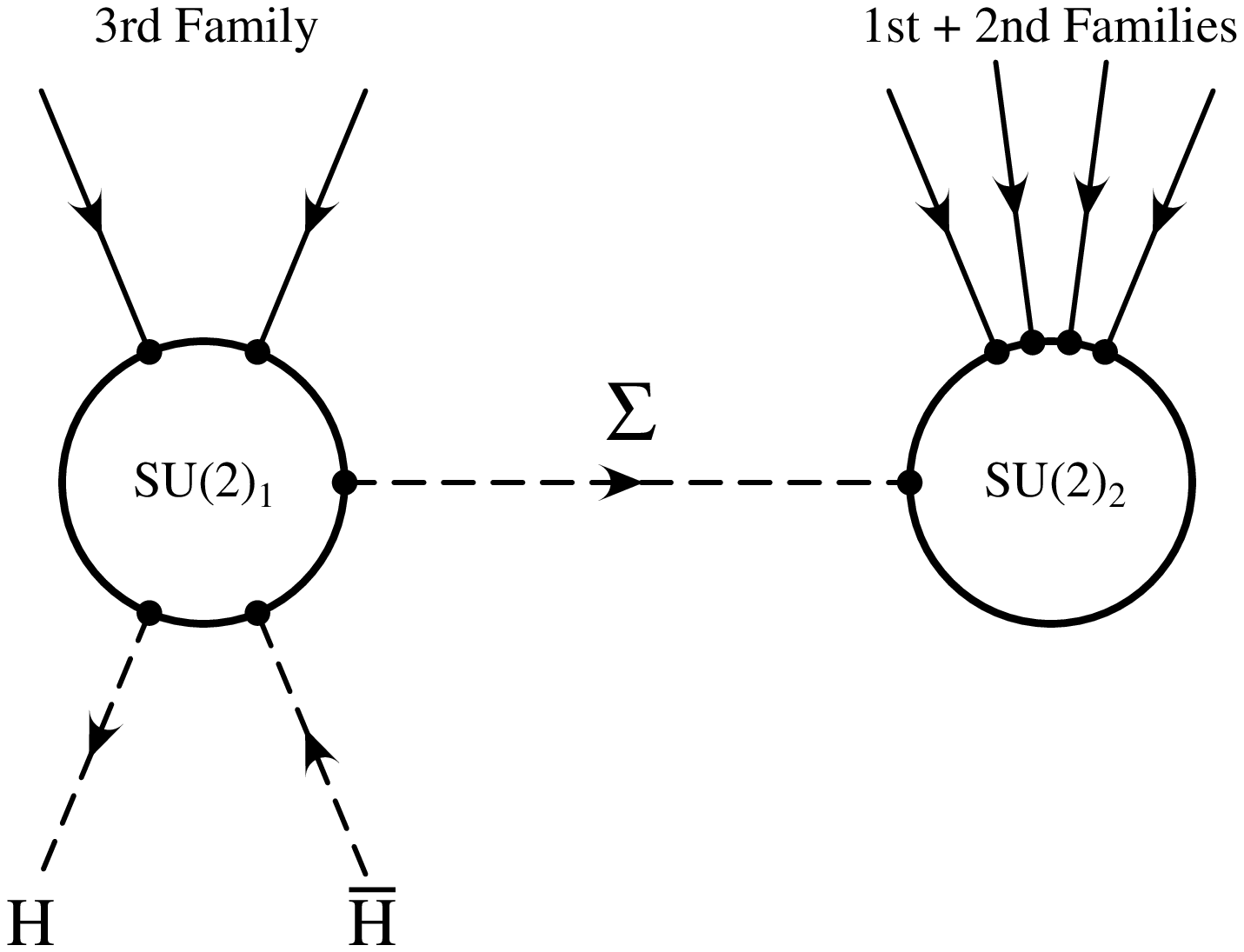, width=3.0in} 
\label{fig:moose}\caption{SU(2) structure}}  
  
Our theory has gauge group  
$SU(3)_c \times SU(2)_1 \times SU(2)_2 \times U(1)_Y$  
with couplings $g_3, g_1, g_2,{\rm and}\  g_Y$,   
respectively. The non-universal charge assignment of the model is  
similar to that of  
Topflavor or ``heavy'' extended technicolor models   
\cite{Chivukula:1995gu} which have $SU(2)_1$ as the  
weak-group for the Higgs fields ($H, \bar{H}$) and third generation of   
fermions, but $SU(2)_2$ as the weak-group for the first two   
generations of fermions. To obtain the correct low energy quantum numbers  
after $SU(2)_1 \times SU(2)_2 \rightarrow SU(2)_D$, the   
third generation and Higgs fields are not charged under $SU(2)_2$,  
and the first two generations are not charged under $SU(2)_1$.   
 
To break the $SU(2)_1 \times SU(2)_2$ to the diagonal subgroup we include   
a bi-doublet chiral field $\Sigma$ which transforms as a   
$(2, \bar{2})$ and gets a VEV ($u$) from superpotential interactions   
with a singlet field, $S_{\Sigma}$. This is depicted schematically in Figure  
\ref{fig:moose}.   
In \cite{Batra:2003nj}, it was shown that the quartic potential for the  
Higgs is enhanced by the additional $SU(2)$ $D$-terms 
when the diagonal symmetry breaking occurs at a scale significantly smaller  
than $m_{\Sigma}$, the soft-mass for $\Sigma$. For this paper, we assume  
that $SU(2)_1 \times SU(2)_2$ breaks to the diagonal $SU(2)$ close to 
the supersymmetric limit, and the enhancement of  
\cite{Batra:2003nj} is negligible. Precision electroweak  
constraints were examined in \cite{Batra:2003nj} 
and require that the breaking scale, $u$, for $SU(2)_1 \times SU(2)_2 
\rightarrow SU(2)_D$ must be $\gtrsim 2.5$ TeV.  We will thus 
consider $u=3$ TeV as an example choice for the symmetry-breaking 
scale for the remainder of this work. The interactions and fields 
needed for Yukawa couplings and diagonal $SU(5)$ unification are 
described below. 
 
To avoid the LEP-II bound without a need for extremely heavy stops, 
we instead enhance the quartic coupling by including a singlet, 
$S$, with superpotential, 
\bea  
\label{eq:W} 
W & = & \mu H \bar{H} + \lambda_{S} S H \bar{H} + \frac{\mu_S}{2}S^2 ,
\eea 
where we self-consistently ignore linear and trilinear terms.  We
also include a soft-mass for $S$, $m_S^2$. We here ignore the effects
of both a soft-trilinear term $A_S S H \bar{H}$ and a Supergravity
generated soft-tadpole $t_s S$. The effect below $m_S$ of $A_S$ would
be to correct the effective $|\lambda_S|^2$ in Eq.~\ref{eq:delta} by
$|A_S|^2 / m_S^2$; the soft-tadpole would induce a VEV for $S$ of
order $t_s / m_S^2$ and could be controlled through low-scale SUSY mediation \cite{Nilles:1997me} or by the appearance of a new gauge group (under which S is charged) at an intermediate scale.

 If both 
$\mu_S \; \& \; m_S \sim v=174 \ {\rm GeV}$, the theory is similar to 
the NMSSM, where $S$ must be included in the Higgs potential and a VEV 
for $S$ could possibly replace the conventional MSSM $\mu$ term.  This 
is an interesting case, but it typically does not realize very large 
increases in the Higgs mass and thus is tightly constrained by the 
LEP-II bound (though small regions of parameter space with in which 
the light state is mostly singlet and thus has suppressed couplings to 
the $Z$ survive) \cite{Miller:2003ay}.  For $\mu_S \gg m_S \& v$, $S$ 
decouples supersymmetrically, and the low energy theory is simply the 
MSSM, including the Higgs sector.  We focus on the third possibility, 
in which $S$ is integrated out in the non-supersymmetric limit, where 
$m_s \gg v$ and $\mu_S^2/(m^2_s+\mu^2_s) \ll 1$.  In 
practice, $\mu_S \sim 100 \ {\rm GeV}$ and $m_S \sim 1~{\rm TeV}$ 
suffice. This is the non-decoupling limit described in 
\cite{Espinosa:1998re} and the limit pursued throughout the rest of 
this paper. 
 
In this limit there are no singlet Higgses in the low energy spectrum,  
and the remnant effect is in the quartic potential for $H$ and $\bar{H}$: 
\bea  
 \frac{g^2}{8} \left(H^{\dagger} 
\vec{\sigma} H- \overline{H} \vec{\sigma} \overline{H}^\dagger 
\right)^2 + \frac{g_Y^2}{8} \left( |\overline{H}|^2-|H|^2 \right)^2 
+\left| \lambda_S \right| ^2 \left| H \bar{H} \right|^2, 
\label{eq:delta}  
\end{eqnarray}  
where $1/g^2=1/g_1^2+1/g_2^2$ is the low energy $SU(2)$ coupling of the 
standard electroweak theory.  The first two terms are the ordinary MSSM  
$D$-terms for the $SU(2) \times U(1)$ gauge interactions.  The last term is  
the effective contribution from $S$ when $\mu_S^2/(m^2_s +\mu^2_s) \ll 1$.  
The Higgs scalar potential also contains the usual MSSM quadratic pieces  
\be  
\left( \left| \mu \right|^2 + m^2_H \right) \left|H \right|^2 +   
\left( \left| \mu \right|^2   
+ m^2_{\bar{H}} \right) \left|\bar{H} \right|^2   
- \left( b H \bar{H} + c.c. \right)  
\ee  
which contribute to electroweak symmetry breaking.   
Throughout, we define $\langle H \rangle = v_H$,   
$\langle \bar{H} \rangle = v_{\bar{H}}$, $v_H^2 + v_{\bar{H}}^2 = v^2$,   
$v=174$ GeV, and $\tan{\beta}= v_{\bar{H}}/v_H$. 
  
General CP-conserving two higgs doublets models were studied in  
\cite{Gunion:2002zf}.   
Unlike the MSSM, this Higgs potential  
has no flat (or negative) $D$-term directions. Electroweak   
symmetry breaking occurs so long as   
$b^2 >(|\mu|^2+m^2_H)(|\mu|^2+m^2_{\bar{H}})$ (which insures   
at least one negative mass eigenvalue in the Higgs mass matrix)   
and $b/\sin{2 \beta} +M_W^2 > v^2 \lambda_S^2$   
(so that $H^{\pm}$ do not develop VEV's and $U(1)_{EM}$ remains unbroken).   
 
\subsection{Yukawa Couplings}  
  
As noted in \cite{Batra:2003nj},   
Yukawa interactions for the third generation fermions  
may be written down at the renormalizable level, since the Higgses and the  
third generation are charged under the same $SU(2)$.  
Yukawa couplings for the first two generations can be generated by adding  
a massive Higgs-like pair of doublets   
$\overline{H}^\prime,H^\prime$, that are charged under $SU(2)_2$. They couple  
to the first two generations via Yukawa-type  
couplings and mix with the regular Higgses via superpotential operators  
such as $\lambda^\prime \overline{H} \Sigma H^\prime$ with $\lambda^{\prime} \langle \Sigma \rangle \sim \mu$.  A supersymmetric mass  
$\mu_{H^\prime} H^\prime \bar{H}^\prime$, with 
$\mu_{H^\prime} \gtrsim \langle\Sigma\rangle$, for the new doublets allows us to  
integrate them out and generates Yukawa couplings  
for the first two generations at low energies.  Mixing with the third generation ({\it i.e.},  
$V_{cb}$ and $V_{ub}$) can easily be generated since the right-handed quarks 
have the same $SU(3)\times U(1)$ quantum numbers.  The result is that the MSSM  
Higgses have essentially MSSM Yukawa interactions, though low $\tan \beta$  
is now accessible as explained in section \ref{sec:constraints}. 
 
\subsection{Grand Unification} 
 
This model can also be made consistent with gauge coupling unification.  
The details are relevant because they influence the $\beta$ function 
coefficients for the gauge couplings---  
important when we determine the bound on $\lambda_S$ by requiring 
perturbativity to the GUT scale in Sec.~\ref{sec:constraints} below. 
The full group $SU(3)_c \times SU(2)_1 \times SU(2)_2 \times U(1)_Y$  
can be embedded in $SU(5)\times SU(5)$ broken by a bi-fundamental field at the GUT scale  
with a VEV $\langle\Xi\rangle = diag\{M,M,M,0,0\}$ \cite{Kribs:2002ew}.  Gauge coupling  
unification  
is predicted (with theoretical uncertainty beyond one-loop) because the  
standard model gauge couplings are only a function of the diagonal gauge  
coupling.  At one loop, one can track the diagonal $SU(2)$ through its  
beta-function coefficient 
as it is the sum of those of the two $SU(2)_i$.  
It receives an extra -6 from the additional triplet of gauge bosons.   
We include a bi-doublet   
$\Sigma$ and two triplets charged under $SU(2)_2$ which altogether  
contribute +6 to the diagonal beta function.  
We must also add an additional vector-like pair of  
triplets to effectively complete a $5$ and $\overline{5}$ with the  
extra pair of Higgs-like fields ($H^\prime$ and $\bar{H}^\prime$) 
required for Yukawa interactions for the first two generations. 
With these additions, both $SU(2)$ models achieve  
the same level of unification as in the MSSM at one loop.  
 
\subsection{Supersymmetry Breaking} 
 
Although our framework is independent of the mechanism of supersymmetry breaking, we do ask that the soft-mass for the singlet be somewhat larger than the soft-masses for the MSSM fields. In gauge mediation, supersymmetry is broken in a hidden sector and communicated 
to the MSSM at one-loop (for gaugino masses) or two-loops (for scalar 
masses) through messenger fields.  The singlet can be coupled directly 
to SUSY breaking messengers $M, \bar{M}$ via the superpotential coupling 
$S M \bar{M}$.  The result is a one-loop (instead of two loop)  
squared soft-mass for $S$ which is 
roughly $4 \pi / \alpha$ larger than the squared MSSM soft-masses, as desired. 
 
Another option which makes use of extra dimensions is gaugino mediation 
\cite{Kaplan:1999ac}: SUSY breaks on a sequestered brane, couples 
directly to all bulk fields (including the gauginos), and then 
communicates to the MSSM matter on a visible brane through the gauginos. 
Gaugino masses arise from a superpotential term of the form $X {\cal 
W}_\alpha {\cal W}^\alpha$, while bulk scalar fields receive  {\it squared}
masses from the K\"ahler term 
$X^\dagger X S^\dagger S$. Therefore, if we put the singlet in the  
bulk, its mass receives an enhancement of $\sqrt{ML}$  
relative to the gaugino masses.  Here, $L$ is the length of the  
extra-dimension, while $M$ is some fundamental scale. Rough bounds from flavor constraints and naive dimensional analysis predict $10 \lesssim ML \lesssim 100$, resulting in the proper enhancement for the soft-mass of $S$.

\section{Perturbativity Constraints}  
\label{sec:constraints} 
 
\FIGURE[t]{ 
\epsfig{file=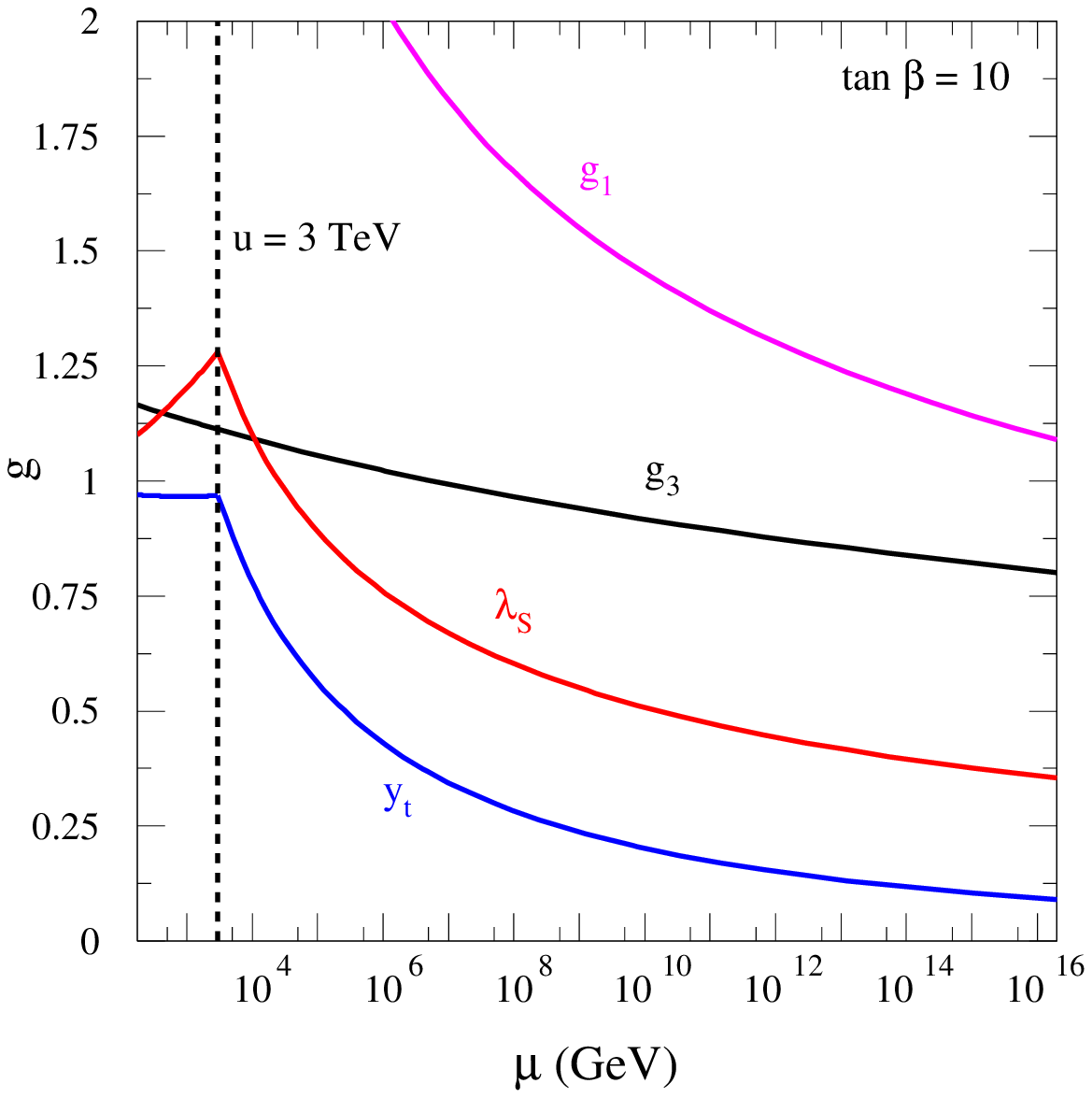,width=.8\linewidth} 
\caption{Renormalization group flow of couplings for the model,  
with $g_1(3\  {\rm TeV}) = 2$,  
$\lambda_S(200 \ {\rm GeV}) = 0.8$  
and $\tan \beta = 10$.}  
\label{fig:rge}} 
 
The enhanced quartic effect in the non-decoupling limit is strongly 
limited by a desire for perturbative unification.  Both couplings 
$\lambda_S$ and $y_t$ feed into each other's renormalization group equations 
(RGE's) with positive coefficients.  
If either $\lambda_S$ or $y_t$ is large at a low scale 
(required for $m_{h^0} > M_Z$, or low $\tan{\beta}$, respectively), 
non-perturbative physics is reached long before $M_{GUT}$. 
 
\FIGURE[t]{ 
\epsfig{file=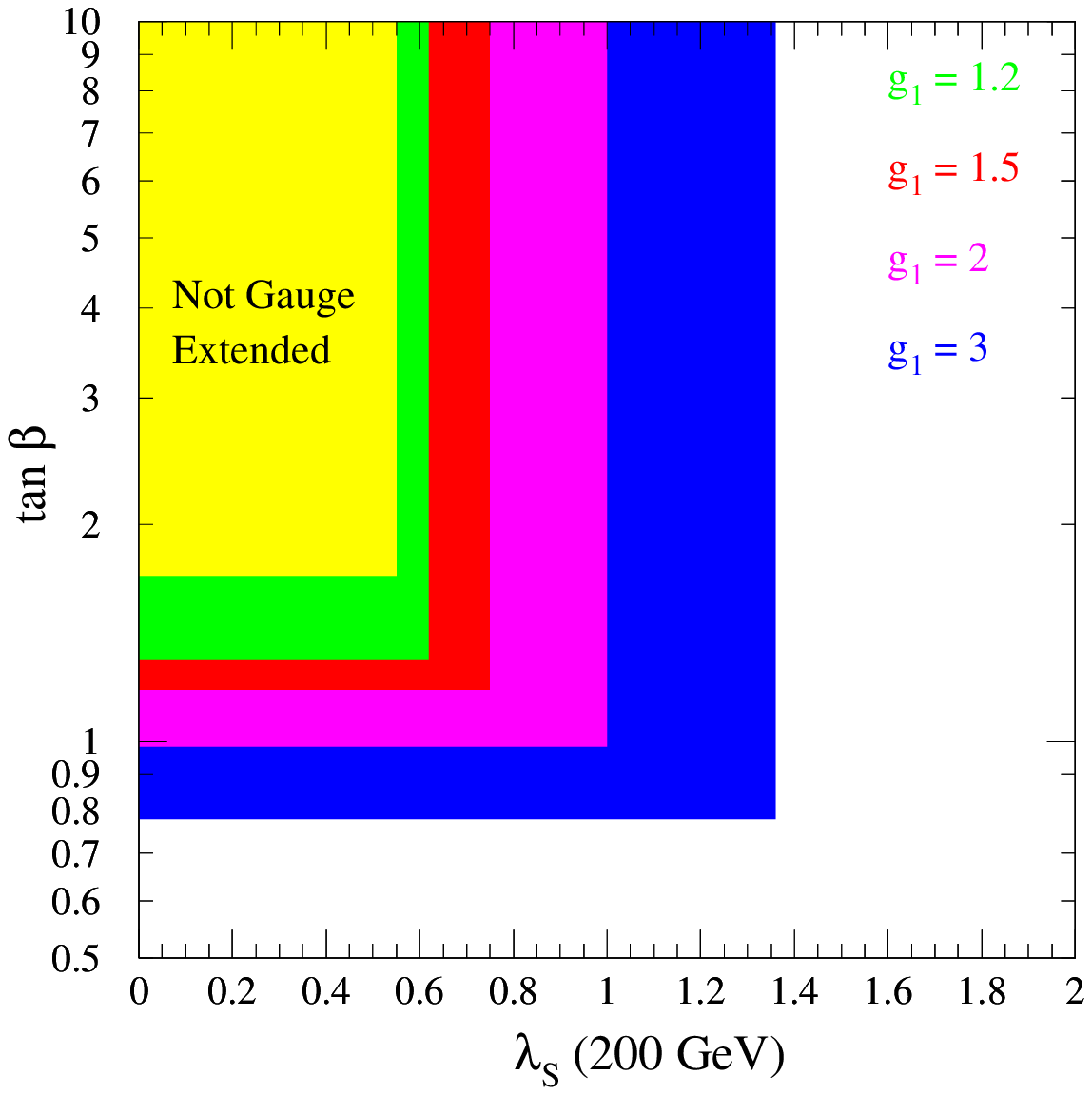,width=.8\linewidth} 
\caption{Allowed regions of $\tan{\beta}$ and $\lambda_S(200 \ {\rm GeV})$  
for given values of the new $SU(2)$ coupling $g_1(3 \ {\rm TeV})$. For the indicated regions, all couplings in the model remain perturbative during 1-loop RG evolution up to the GUT scale.  From left 
to right, the regions are: the non-gauge extended model of 
\protect{\cite{Espinosa:1998re}}, and our gauge extension with 
$g_1(u)=$1.2, 1.5, 2, and 3.} 
\label{fig:tanbeta}} 
  
Both of these problems are largely ameliorated by the presence of new, relatively strong,  
gauge interactions, which drive both $y_t$ and $\lambda_S$   
down at large scales, owing to the Higgs and top quark participation in the  
stronger group.  The dominant terms in the renormalization group equations 
at one loop (including the spectator matter described above, necessary for 
gauge coupling unification and Yukawa interactions for the first two  
generations) are\footnote{These RGE's are valid above the  
$SU(2) \times SU(2)$ breaking scale, $u$.  Below $u$, we use the RGE's 
appropriate for the broken phase.}, 
\bea 
\frac{d g_1}{dt}&=&-2 \frac{ g_1^3}{16 \pi^2} \nonumber \\   
\frac{d g_2}{dt}&=&4 \frac{ g_2^3}{16 \pi^2} \nonumber \\   
\frac{d g_3}{dt}&=&-2 \frac{ g_3^3}{16 \pi^2} \nonumber \\  
\frac{d \lambda_S}{dt}&=&\frac{\lambda_S}{16 \pi^2}  
\left( 3 \left| y_t \right|^2 + 4\left| \lambda_S \right|^2  
- 3 g_1^2 \right) \nonumber \\  
\frac{d y_t}{dt}&=&\frac{y_t}{16 \pi^2}\left( 6 \left| y_t \right|^2  
+  \left| \lambda_S \right|^2 - 3g_1^2-\frac{16}{3}g_3^2 \right)   , 
\eea 
with sub-dominant contributions from the bottom Yukawa 
(particularly at low $\tan \beta$), and $g_Y$. 
For large enough $g_1$, the strong gauge interactions drive $y_t$ and 
$\lambda_S$ smaller just above the electroweak scale.   
The effectiveness is limited by the fact that the   
additional $SU(2)$ is itself strongly asymptotically-free, and thus becomes  
more and more irrelevant in the RGEs at higher and higher energies.  However, 
the result is a wider region of allowed parameter space consistent with 
perturbativity to the GUT scale. 
In Figure~\ref{fig:rge} we show a sample flow of the couplings for 
$\tan \beta = 10$, $g_1(u = 3~{\rm TeV}) = 2$, and  
$\lambda_S(200~{\rm GeV}) = 0.8$. 
 
In Figure~\ref{fig:tanbeta} we show the allowed regions of $\tan \beta$ and 
$\lambda_S$, for fixed values of $g_1(u=3 ~{\rm TeV})$, by requiring that 
all couplings remain perturbative up to the GUT scale.   
It should be noted that 
there is in fact a minimum value of $g_1$ for this theory which is compatible 
with perturbative unification.  Because $g_2$ is not asymptotically-free, if 
its value at $u$ is too large, it will flow strong before the GUT scale. 
From the one-loop RGE expression above, we see that this happens when 
$g_1 \leq 1.2$ at $u=3$ TeV. 
  
\section{Higgs Properties and Phenomenology}  
\label{sec:pheno} 
  
\subsection{Higgs Spectrum} 
 
We include stop corrections and find that the one-loop Higgs CP-even mass matrix can be written as 
a function of the CP-odd mass ($m_A$), the $\mu$ parameter, the stop 
masses $m_{\tilde{t_i}}$, and the stop mixing parameter, $A_t$ 
\cite{Ellwanger:1996gw}:  
 
\bea m_{11}^2&=&m_Z^2 
\cos^2\beta+m_A^2\sin^2\beta-\frac{3 y_t^2} 
{16\pi^2}\mu^2\frac{Z^2}{3}\nonumber\\ m_{22}^2&=&m_Z^2 
\sin^2\beta+m_A^2\cos^2\beta+\frac{3 y_t^2} {16\pi^2}\left(4 m_t^2 
\log\frac{m_{\tilde{t_1}} m_{\tilde{t_2}}}{m_t^2}+A_t(2 m_t Z- 
A_t\frac{Z^2}{3})\right) \nonumber\\ m_{12}^2&=&-\frac{1}{2} 
(m_Z^2+m_A^2-2v^2\lambda_S^2)\sin 2 \beta+\frac{3 y_t^2} {16\pi^2}\mu 
\left(m_t Z-A_t\frac{Z^2}{3}\right) 
\label{one-loop}  
\eea  
where:  
\be  
Z=\frac{m_t(A_t+\mu\cot\beta)}{m_t^2+\frac{1}{2}(m^2_Q+m^2_U)}  
\ee  
and the stop masses are defined with respect   
to the soft-masses for left- and right-handed stops ($m_Q$,$m_U$) as:  
\bea  
m^2_{\tilde{t}_{1,2}}&=&m_t^2+\frac{1}{2}(m^2_Q+m^2_U)\pm W\nonumber\\  
W^2&=&\frac{1}{4}(m^2_Q-m^2_U)^2+y_t^2 v^2 |A_t \sin\beta-\mu \cos\beta|^2  
\eea  
 
The charged Higgs mass is (at one-loop): 
\bea  
m^2_{H^{\pm}}= m^2_{A}+m^2_W - v^2 \left|\lambda_S\right|^2.  
\eea 
 
In the above, the 
top Yukawa coupling is evaluated at the stop mass scale in order to  
take into account the leading effect from RGE-improvement \cite{Carena:1995wu}. 
We have neglected the sub-dominant corrections from the gauginos and 
superpartners of light fermions. 
The CP-even mass eigenstates and mixing angle $\alpha$ are obtained by 
diagonalizing this two by two matrix.  The largest $h^0$ masses are obtained 
for large $\lambda_S$, the decoupling regime $m_A > m_{h^0}$, and 
$\tan \beta \sim 1$.  For $\lambda_S \sim 1.5$, these parameters  
(with soft-masses for the stops of 200 GeV) result in 
$m_{h^0} \sim 260$ GeV, which is the largest $h^0$ mass that can be realized 
in our model consistent with perturbativity up to the GUT scale. Such  
a large value of $\lambda_S$ starts to reintroduce fine-tuning into 
the Higgs soft-mass through the renormalization group equations if SUSY 
is broken close to the Planck scale. For $m_S \sim 1\ $ TeV  
and $\lambda_S \sim 1.5$ this contribution is less than $1\ $TeV. 
 
In Fig.~\ref{fig:hmass} we plot the spectrum for $m_Q = m_U = \mu = 200$ GeV, 
$\tan \beta = 1$ and $\lambda_S = 0.8$ (requiring 
$g_1(u) \sim 2$, see Fig.~\ref{fig:tanbeta}), as a function of $m_A$. 
It should be noted that this set of soft SUSY-breaking 
parameters in the MSSM would predict a lightest CP-even Higgs mass much below  
the LEP-II bound, and would thus be ruled out.  
We see that even for these rather small soft-masses  
for the stops (and no stop mixing at all), 
we can easily accommodate CP-even Higgs masses consistent with the LEP-II 
bounds.  This eliminates the fine-tuning of the weak scale that 
is inevitable in the MSSM. 

\FIGURE[t]{ 
\epsfig{file=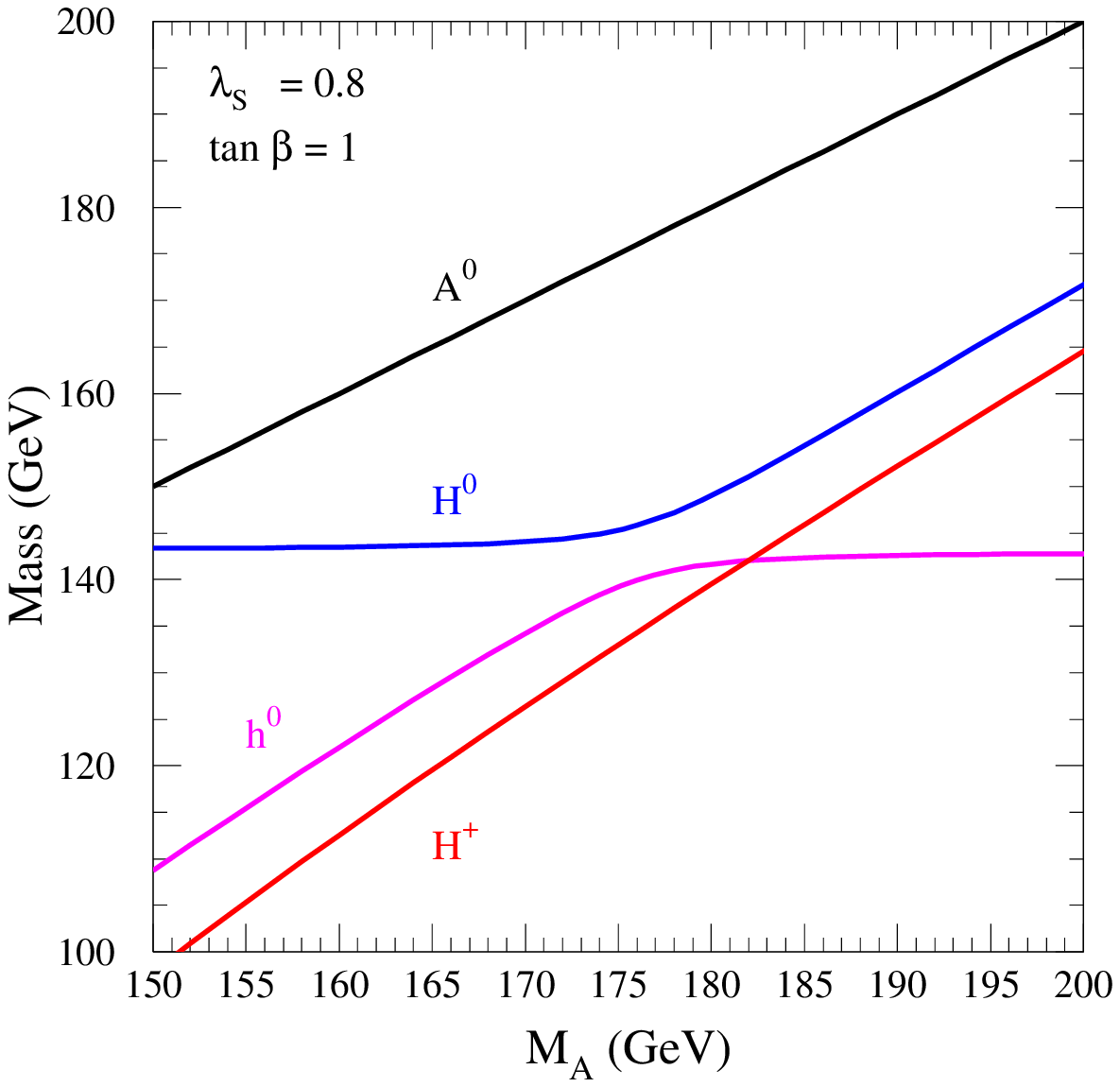,width=.8\linewidth} 
\caption{One loop spectrum of Higgs masses (varying $m_A$)  
for $\tan\beta=1$ and $\lambda_S=0.8$.  The soft parameters were chosen 
as $m_Q = m_U = \mu = 200$ GeV and $A_t=0$.} 
\label{fig:hmass}} 
 
\FIGURE[t]{ 
\epsfig{file=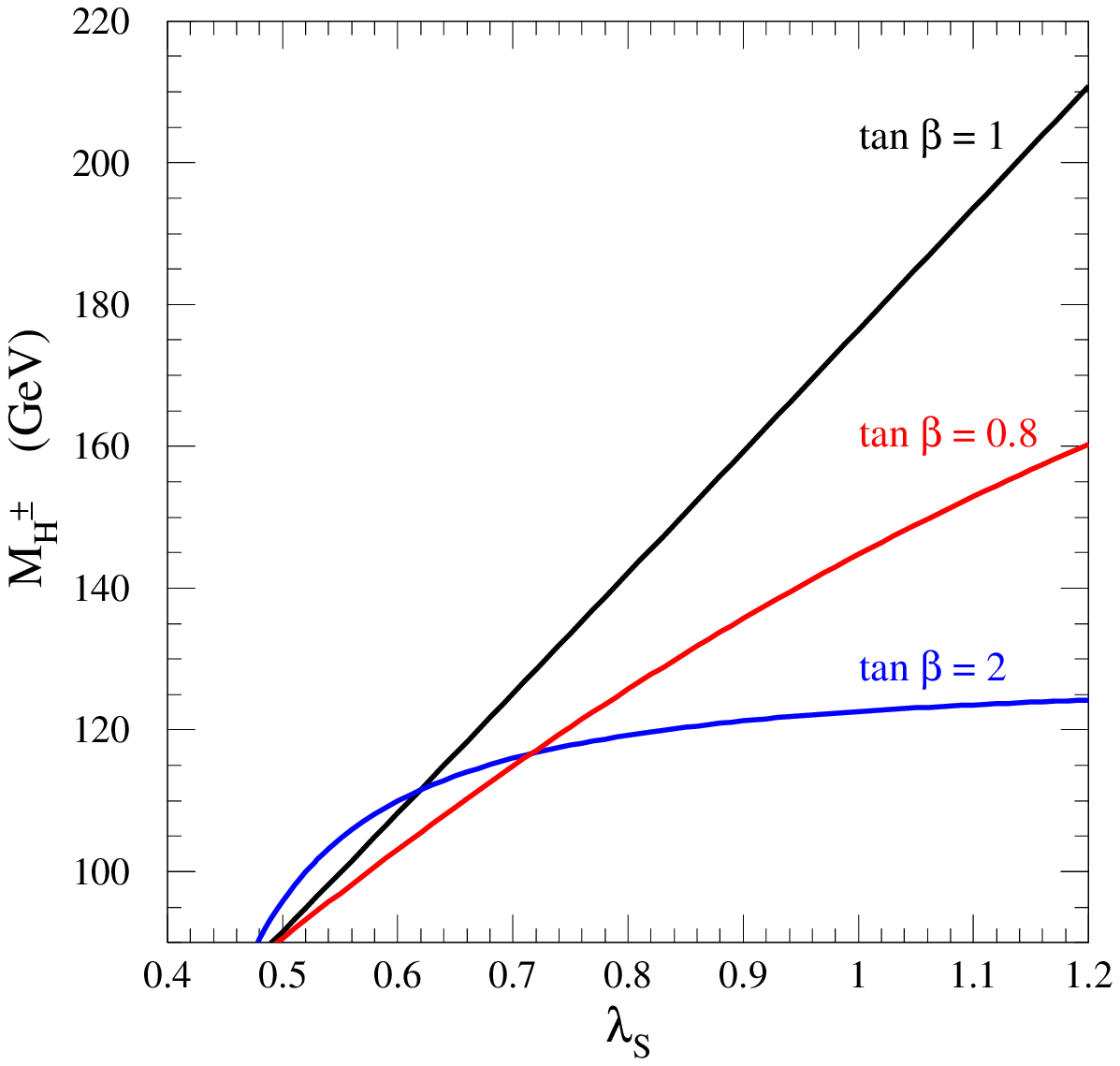,width=.8\linewidth} 
\caption{The charged Higgs mass versus $\lambda_S$ for different values of $\tan{\beta}$.
The area below each line represents the parameter space in which the charged Higgs is the
lightest Higgs in the spectrum.  The lines themselves represent the charged Higgs masses for which the charged and lightest CP-even Higgs bosons are degenerate.  The soft parameters were chosen as $m_Q = m_U = \mu = 200$ GeV and $A_t=200$ GeV, and there is moderate dependence on them.  For example, if the above parameters are set at 500 GeV, the values
of the lines at $\lambda=1.2$ are (from lowest to highest $\tan{\beta}$)  162, 210, and 145 GeV. } 
\label{fig:chghiggs}}

\subsection{Collider Phenomenology}  
 
This theory has a rich and distinct Higgs phenomenology compared to
the MSSM.  For example, Figure~\ref{fig:hmass} shows a sample Higgs
spectrum for $\tan{\beta}=1$. Figure~\ref{fig:chghiggs} indicates the
region of parameter space where the lightest state is the
\emph{charged higgs} corresponding to the case when the heavier Higgs
is responsible for electroweak symmetry-breaking.  

The LEP-II limits on the charged Higgs mass (from direct searches such
as $e^+ e^- \rightarrow H^+ H^-$) is 81 GeV \cite{lep2charged} and
requires $m_A > 140$ GeV if $\lambda_S = 0.8$ and $\tan \beta = 1$.
In this regime the light CP-even Higgs has small couplings to gauge
bosons and may avoid LEP-II bounds even for $m_{h} \leq 115$ GeV.
Indirect constraints on $m_{H^+}$ from the observed rate of
$b\rightarrow s \gamma$ (see \cite{Ali:2002jg} for a recent summary)
are stringent even in models with minimal flavor violation and
superpartners at the TeV scale. However, supersymmetric cancellations
from loops involving light charginos and stops (see
\cite{Barbieri:1993av}, for example) can reproduce the observed $b
\rightarrow s \gamma $ rates even for $m_{H^+} \sim 80\ {\rm GeV}$ and
$\tan{\beta} \sim 1$ \cite{Ciuchini:1998xy}.  For $m_{H^+} \sim 80 \
{\rm GeV}$ and $\tan{\beta} \sim 1$, the $m_{H^+}$ contribution is
cancelled by $\sim 75 \%$.

For much of the region of parameters in which it is the lightest, the charged 
Higgs mass is much less than the top mass.   
This implies (for low $\tan \beta$) that the charged Higgs will 
predominantly decay into charm and strange, with a much smaller fraction into 
$\tau$ and neutrino. 
For $m_{H^\pm} \sim m_t$ the three-body decay through an off-shell top 
into $W b \bar{b}$ becomes interesting. 
The dominant production for this range of masses 
is through a rare top decay, $t \rightarrow H^+ b$. 
The branching fraction is given at tree-level by, 
\bea 
\frac{\Gamma (t \rightarrow H^+ b)}{\Gamma (t \rightarrow W^+ b)} & = & 
\frac{1}{1+2 m_W^2/m_t^2} \left( \frac{1 - m_{H^\pm}^2 / m_t^2}{1 - m_W^2 / m_t^2} 
\right)^2 \times \cot^2 \beta 
\eea 
and is enhanced for $\tan \beta < 1$. 
One-loop SUSY QCD corrections tend to suppress the branching ratio 
slightly \cite{Carena:1999py}. 
This decay can be distinguished from usual top decay first because it modifies 
the branching ratios into jets compared to leptons, and second because the 
decay products reconstruct an intermediate $H^\pm$ mass instead of the $W$  
mass.  At the Tevatron run II, predictions are that the region with 
$\tan \beta < 1$ and $m_{H^\pm} < 120$ GeV can be discovered or excluded 
through top decay  
with 2 fb$^{-1}$ of integrated luminosity \cite{Carena:2000yx}.  The LHC 
with 100 fb$^{-1}$ 
is expected to be able to to see a charged Higgs with $m_{H^\pm} < m_t - 20$ 
GeV for all values of $\tan \beta$ \cite{Beneke:2000hk}. 
 
Charged Higgs masses greater than $m_t$ are possible  
(continuing to assume $\lambda_S = 0.8$) if $m_A$ is larger than 210 GeV. 
In this case, the dominant decay is $H^+ \rightarrow t \overline{b}$ and 
the dominant production at the LHC is associated production of a Higgs with 
a top quark, through partonic processes such as $g b \rightarrow H^- t$ 
\cite{Barnett:1987jw}.  This process can study charged Higgs bosons with masses 
up to about 400 GeV in the low $\tan \beta$ region \cite{Cavalli:2002vs} 
with 100  fb$^{-1}$.  Another production mechanism is through an off-shell 
$W$ boson, $q \bar{q}^\prime \rightarrow W^* \rightarrow H^+ A_0$  
\cite{Cao:2003tr}, leading to final states  
with $t \bar{b} b \bar{b}$.  Two of the bottoms  
reconstruct $m_{A^0}$, and thus typically have much higher energies than 
$b \bar{b}$ from gluon splitting. 
 
Note that despite the absence of additional weak scale Higgs bosons, this 
theory modifies the usual MSSM $m^2_{H^\pm}-m_A^2 = m^2_W$  
mass relation.  Even when 
the spectrum is roughly consistent with the MSSM (say, for modest  
$\lambda_S$ and large $m_A$ such that $m_{h^0}$ has mass just above 
the LEP bound and $h^0$ has largely SM-like couplings),  
this fact can be tested at the 
LHC through the associated production of $H^\pm A^0$ provided  
the masses are less than about 300 GeV \cite{Cao:2003tr}. 
As with any extension of the MSSM that affects the Higgs quartic,  
one could also combine a measurement of the light CP even Higgs mass with  
precision measurements of the stop masses and mixing angle (for example, 
at a linear collider \cite{Bartl:2000kw})  
to show that the Higgs mass does not satisfy the MSSM relation. 
 
The additional gauge bosons (with masses in the several TeV range) 
associated with the top-flavor group can be 
produced at the LHC.  The dominant process is one in which two first generation 
quarks fuse into a $W^\prime$.  The $W^\prime$ coupling to light quarks 
is suppressed compared to third generation fermions,  
but this is compensated by the much larger probability to find a  
high-energy first generation (valence) quark inside a proton 
compared to the probability of finding a third generation quark. 
Once 
produced, the $W^\prime$ predominantly decays into $t \overline{b}$ and 
$\tau \overline{\nu}$, and their super-partners.  The decay into  
$t \overline{b}$ looks like $s$-channel single top production 
\cite{Tait:2000sh} and has been carefully studied in  
Ref.~\cite{Sullivan:2003xy}, including NLO QCD  
corrections to the signal rate, estimations of detector efficiencies and 
backgrounds, etc.  The conclusion of that study is that $W^\prime$ masses 
less than 4.5 TeV can discovered with 100 fb$^{-1}$ of integrated luminosity. 
 
Finally, the additional gauge fields and the $\Sigma$ 
provide a number of heavy fermions with masses around $u \sim 3$ TeV, 
and the super-partner of the singlet $S$ is typically weak scale. 
They will have small (but potentially important) mixings with the standard 
charginos and neutralinos, and strong mixing with each other.  The gaugino 
components will also prefer to couple to the third family, and decays 
involving multiple bottom and top quarks (plus missing energy from the LSP) 
and $\tau$ leptons will dominate.  The masses are probably too large for  
copious production at the 
LHC, but the influence through mixing on the lighter neutralinos and charginos 
could affect their production and decay in a relevant way.  Precision  
measurements of neutralino and chargino couplings  
at a future $e^+ e^-$ linear collider could reveal this mixing. 
 
\section{Conclusions}  
\label{sec:conclusions} 
  
The need for a Higgs mass greater than the LEP-II bound has placed the 
MSSM in an interesting position 
and motivates extensions which allow for larger Higgs masses. 
In this article we have presented one such model which invokes a 
singlet Higgs to increase the Higgs mass by effectively adding a new Higgs 
quartic to the superpotential.  The new feature is the addition of 
asymptotically-free gauge interactions which tend to drive this extra 
quartic and the top Yukawa coupling smaller at high 
energies. 
 
This allows one to explore larger quartic couplings consistent with 
perturbative unification than in the past \cite{Espinosa:1998re} and predicts 
a new bound on the lightest CP-even state $m_{h^0} < 250 \ {\rm GeV}$. 
It also opens a window of $\tan \beta < 1$ for which the top Yukawa 
remains perturbative all the way up to the GUT scale.  The result is a 
theory in which the lightest CP-even Higgs may be heavier than the 
LEP-II bound at tree level without the need to invoke large stop masses 
and introduce electroweak fine-tuning.  In fact, the phenomenology associated 
with larger quartics and lower $\tan \beta$ is more general than the specific 
model presented here; a theory such as the ``SUSY Fat Higgs'' model  
\cite{Harnik:2003rs} which 
invokes a low scale cut-off on $\lambda_S$ and the Yukawa interactions 
(while still remaining consistent with unification)  
also allows one to explore the same 
regions of parameter space with similar phenomenology. 
 
 
Replacing the singlet with a pairs of triplets (with hypercharge $\pm$1) as in  
Ref~\cite{Espinosa:1998re} is unlikely to enhance the mass-bound further  
as the additional matter will make the $\beta$-function of $g_1$ 
vanish at one-loop.  Further, a triplet contribution to the Higgs quartic 
will inevitably lead to a VEV for the triplet, which is disfavored by precision 
electroweak data. 
Instead one could imagine enhancing the bound by additionally including  
the mechanism of 
Ref.~\cite{Batra:2003nj} by lifting the soft-mass for the gauge-breaking field 
to $\sim$ 10 TeV. 
 
The resulting phenomenology is 
somewhat unusual with the charged Higgs as the lightest Higgs for a 
range of parameters.  The additional gauge bosons are also expected to be 
visible at the LHC and provide a tangible way in which experiments 
would be able to test this scenario and distinguish it from 
alternatives.

\acknowledgments 
 
The authors have benefited from discussions with 
M. Carena, S. Mrenna, E. Ponton, S. Thomas, J. Wacker,
C.E.M. Wagner, N. Weiner, and C.P. Yuan.  A.D., D.E.K. and P.B. are  
supported by NSF Grants P420D3620414350 and P420D3620434350. 
Fermilab is operated by Universities Research Association Inc.  under  
contract no. DE-AC02-76CH02000 with the DOE.


\begin{thebibliography}{99}  
 
 
\bibitem{unknown:2001xx} 
  [LEP Higgs Working Group Collaboration], 
arXiv:hep-ex/0107030. 
 
\bibitem{Carena:1995wu} 
Y.~Okada, M.~Yamaguchi and T.~Yanagida, 
Prog.\ Theor.\ Phys.\  {\bf 85}, 1 (1991); 
M.~Carena, M.~Quiros, and C.~E.~M.~Wagner, 
Nucl.\ Phys.\ B {\bf 461}, 407 (1996) 
[arXiv:hep-ph/9508343]; 
H.~E.~Haber, R.~Hempfling, and A.~H.~Hoang, 
Z.\ Phys.\ C {\bf 75}, 539 (1997) 
[arXiv:hep-ph/9609331]; 
J.~R.~Espinosa and I.~Navarro, 
Nucl.\ Phys.\ B {\bf 615}, 82 (2001) 
[arXiv:hep-ph/0104047]. 

 
 
 
 
 
\bibitem{Ellis:1988er} 
J.~R.~Ellis, J.~F.~Gunion, H.~E.~Haber, L.~Roszkowski and F.~Zwirner, 
Phys.\ Rev.\ D {\bf 39}, 844 (1989); 
H.~E.~Haber and M.~Sher, 
Phys.\ Rev.\ D {\bf 35}, 2206 (1987); 
M.~Drees, 
Phys.\ Rev.\ D {\bf 35}, 2910 (1987); 
K.~S.~Babu, X.~G.~He and E.~Ma, 
Phys.\ Rev.\ D {\bf 36}, 878 (1987); 
M.~Cvetic, D.~A.~Demir, J.~R.~Espinosa, L.~L.~Everett and P.~Langacker, 
Phys.\ Rev.\ D {\bf 56}, 2861 (1997) 
[Erratum-ibid.\ D {\bf 58}, 119905 (1998)] 
[arXiv:hep-ph/9703317]. 
 
\bibitem{Miller:2003ay} 
D.~J.~Miller, R.~Nevzorov and P.~M.~Zerwas, 
Nucl.\ Phys.\ B {\bf 681}, 3 (2004) 
[arXiv:hep-ph/0304049]; 
U.~Ellwanger, J.~F.~Gunion, C.~Hugonie and S.~Moretti, 
arXiv:hep-ph/0401228. 
 
\bibitem{Haber:1986gz} 
J.~R.~Espinosa and M.~Quiros, 
Phys.\ Lett.\ B {\bf 279}, 92 (1992); 
J.~R.~Espinosa and M.~Quiros, 
Phys.\ Lett.\ B {\bf 302}, 51 (1993) 
[arXiv:hep-ph/9212305]. 
 
\bibitem{Espinosa:1998re} 
J.~R.~Espinosa and M.~Quiros, 
Phys.\ Rev.\ Lett.\  {\bf 81}, 516 (1998) 
[arXiv:hep-ph/9804235]. 
 
\bibitem{Giudice:1988yz} 
G.~F.~Giudice and A.~Masiero, 
Phys.\ Lett.\ B {\bf 206}, 480 (1988); 
L.~J.~Hall and L.~Randall, 
Phys.\ Rev.\ Lett.\  {\bf 65}, 2939 (1990); 
E.~J.~Chun, J.~E.~Kim and H.~P.~Nilles, 
Nucl.\ Phys.\ B {\bf 370}, 105 (1992); 
N.~Arkani-Hamed, L.~J.~Hall, D.~R.~Smith and N.~Weiner, 
Phys.\ Rev.\ D {\bf 63}, 056003 (2001) 
[arXiv:hep-ph/9911421]. 
 
\bibitem{Tobe:2002zj} 
K.~Tobe and J.~D.~Wells, 
Phys.\ Rev.\ D {\bf 66}, 013010 (2002) 
[arXiv:hep-ph/0204196]. 
 
\bibitem{Batra:2003nj} 
P.~Batra, A.~Delgado, D.~E.~Kaplan and T.~M.~P.~Tait, 
JHEP {\bf 0402}, 043 (2004) 
[arXiv:hep-ph/0309149]. 
 
\bibitem{Casas:2003jx} 
J.~A.~Casas, J.~R.~Espinosa and I.~Hidalgo, 
JHEP {\bf 0401}, 008 (2004) 
[arXiv:hep-ph/0310137]. 
 
\bibitem{Harnik:2003rs} 
R.~Harnik, G.~D.~Kribs, D.~T.~Larson and H.~Murayama, 
arXiv:hep-ph/0311349. 
  
\bibitem{Panagiotakopoulos:2000wp}
C.~Panagiotakopoulos and A.~Pilaftsis,
Phys.\ Rev.\ D {\bf 63}, 055003 (2001)
[arXiv:hep-ph/0008268].
C.~Panagiotakopoulos and A.~Pilaftsis,
Phys.\ Lett.\ B {\bf 505}, 184 (2001)
[arXiv:hep-ph/0101266].




\bibitem{Chivukula:1995gu}  
R.~S.~Chivukula, E.~H.~Simmons and J.~Terning,  
Phys.\ Rev.\ D {\bf 53}, 5258 (1996)  
[arXiv:hep-ph/9506427]; 
D.~J.~Muller and S.~Nandi,  
Phys.\ Lett.\ B {\bf 383}, 345 (1996)  
[arXiv:hep-ph/9602390];  
E.~Malkawi, T.~Tait and C.~P.~Yuan,  
Phys.\ Lett.\ B {\bf 385}, 304 (1996)  
[arXiv:hep-ph/9603349]; 
H.~J.~He, T.~Tait and C.~P.~Yuan, 
Phys.\ Rev.\ D {\bf 62}, 011702 (2000) 
[arXiv:hep-ph/9911266]. 
 


 
 
 
 
 
  


\bibitem{Nilles:1997me}
H.~P.~Nilles and N.~Polonsky,
Phys.\ Lett.\ B {\bf 412}, 69 (1997)
[arXiv:hep-ph/9707249].

 
\bibitem{Gunion:2002zf}  
J.~F.~Gunion and H.~E.~Haber,  
Phys.\ Rev.\ D {\bf 67}, 075019 (2003)  
[arXiv:hep-ph/0207010].  
  
\bibitem{Martin:1997ns}  
S.~P.~Martin,  
arXiv:hep-ph/9709356.  
  
\bibitem{Kribs:2002ew}  
G.~D.~Kribs,  
{\it Prepared for 10th International Conference on Supersymmetry and  
Unification of Fundamental Interactions (SUSY02), Hamburg, Germany, 17-23  
Jun 2002}. 
 
\bibitem{Kaplan:1999ac} 
D.~E.~Kaplan, G.~D.~Kribs and M.~Schmaltz, 
Phys.\ Rev.\ D {\bf 62}, 035010 (2000) 
[arXiv:hep-ph/9911293]; 
Z.~Chacko, M.~A.~Luty, A.~E.~Nelson and E.~Ponton, 
JHEP {\bf 0001}, 003 (2000) 
[arXiv:hep-ph/9911323]. 
 
\bibitem{Ellwanger:1996gw}  
U.~Ellwanger, M.~Rausch de Traubenberg and C.~A.~Savoy,  
Nucl.\ Phys.\ B {\bf 492}, 21 (1997)  
[arXiv:hep-ph/9611251].  
 
\bibitem{lep2charged} 
LEP Higgs Working group,  
http://lephiggs.web.cern.ch/LEPHIGGS/pdg2004/index.html. 

\bibitem{Ali:2002jg}
A.~Ali, E.~Lunghi, C.~Greub and G.~Hiller,
Phys.\ Rev.\ D {\bf 66}, 034002 (2002)
[arXiv:hep-ph/0112300].

\bibitem{Barbieri:1993av}
R.~Barbieri and G.~F.~Giudice,
Phys.\ Lett.\ B {\bf 309}, 86 (1993)
[arXiv:hep-ph/9303270].
J.~L.~Hewett and J.~D.~Wells,
Phys.\ Rev.\ D {\bf 55}, 5549 (1997)
[arXiv:hep-ph/9610323].

\bibitem{Ciuchini:1998xy}
M.~Ciuchini, G.~Degrassi, P.~Gambino and G.~F.~Giudice,
Nucl.\ Phys.\ B {\bf 534}, 3 (1998)
[arXiv:hep-ph/9806308].




  
\bibitem{Carena:1999py} 
M.~Carena, D.~Garcia, U.~Nierste and C.~E.~M.~Wagner, 
Nucl.\ Phys.\ B {\bf 577}, 88 (2000) 
[arXiv:hep-ph/9912516]. 
 
\bibitem{Carena:2000yx}  
M.~Carena {\it et al.}  [Higgs Working Group Collaboration],  
arXiv:hep-ph/0010338.  
 
\bibitem{Beneke:2000hk} 
M.~Beneke {\it et al.}, 
arXiv:hep-ph/0003033. 
 
\bibitem{Barnett:1987jw} 
R.~M.~Barnett, H.~E.~Haber and D.~E.~Soper, 
Nucl.\ Phys.\ B {\bf 306}, 697 (1988); 
A.~C.~Bawa, C.~S.~Kim and A.~D.~Martin, 
Z.\ Phys.\ C {\bf 47}, 75 (1990); 
V.~D.~Barger, R.~J.~N.~Phillips and D.~P.~Roy, 
Phys.\ Lett.\ B {\bf 324}, 236 (1994) 
[arXiv:hep-ph/9311372]; 
E.~L.~Berger, T.~Han, J.~Jiang and T.~Plehn, 
arXiv:hep-ph/0312286. 
 
\bibitem{Cavalli:2002vs} 
D.~Cavalli {\it et al.}, 
arXiv:hep-ph/0203056. 
 
\bibitem{Cao:2003tr} 
S.~Kanemura and C.~P.~Yuan, 
Phys.\ Lett.\ B {\bf 530}, 188 (2002) 
[arXiv:hep-ph/0112165]; 
Q.~H.~Cao, S.~Kanemura and C.~P.~Yuan, 
arXiv:hep-ph/0311083. 
 
\bibitem{Bartl:2000kw} 
A.~Bartl, H.~Eberl, S.~Kraml, W.~Majerotto and W.~Porod, 
Eur.\ Phys.\ J.\ directC {\bf 2}, 6 (2000) 
[arXiv:hep-ph/0002115]; 
A.~Finch, H.~Nowak and A.~Sopczak, 
arXiv:hep-ph/0211140; 
E.~L.~Berger, J.~Lee and T.~M.~P.~Tait, 
Phys.\ Rev.\ D {\bf 69}, 055003 (2004) 
[arXiv:hep-ph/0306110]. 
 
\bibitem{Tait:2000sh} 
T.~Tait and C.~P.~Yuan, 
Phys.\ Rev.\ D {\bf 63}, 014018 (2001) 
[arXiv:hep-ph/0007298]. 
 
\bibitem{Sullivan:2003xy} 
Z.~Sullivan, 
arXiv:hep-ph/0306266. 
 
  
 
\end{thebibliography}
\end{document}